\documentclass[paper]{revtex4}

\usepackage{graphicx}
\usepackage{axodraw}
\usepackage{epsfig}
\usepackage{fancyhdr}
\usepackage{empheq}
\usepackage{mathbbol}
\usepackage{wasysym}
\usepackage{pstricks}
\usepackage{color}
\usepackage{bbm}
\usepackage{esint}
\def\qq{\mathbbmss{q}}

\def\aa{{\cal A}}

\begin{document}

\title{A note on large $N$ scalar QCD$_2$}

\author{B Grinstein$^{\dag}$, R Jora$^{ *}$ and  AD Polosa$^*$
} 
\affiliation{$^\dag$University of California, San Diego; Dept. of Physics, La Jolla, CA92093-0315, USA\\
$^*$INFN Roma, Piazzale A Moro 2, Roma, I-00185, Italy}

\begin{abstract}
We review the features of the bound state equation in large $N$ scalar QCD in two dimensions,
the 't~Hooft model, and compute the discrete hadron mass spectrum  in this theory.
We make the Ansatz that the scalar fields of this model represent spin zero diquarks and we estimate the
minimum allowed mass for the first radial excitation of the lowest diquark-antidiquark  scalar meson.
The discussion is extended to the case of spin one diquarks.
\end{abstract}

\maketitle

\thispagestyle{fancy}

{\bf \emph{ Introduction}}.
In a recent paper on scalar meson dynamics~\cite{hooft} it is shown how a satisfactory explanation 
of light scalar meson decays can be reached by assuming a dominant diquark-antidiquark structure,
$\qq\bar \qq$, for the lightest particles.  The diquark $\qq$ is  a spin zero  antitriplet color state.  
In first approximation the nonet formed by $f_0(980),a_0(980),\kappa(900),\sigma(500)$ is
interpreted as the lowest $\qq\bar \qq$ multiplet. On the other hand, the decuplet 
of scalar mesons with masses above  1~GeV, formed by $f_0(1370),f_0(1500),f_0(1710), a_0(1450),K_0(1430)$, and likely
containing the lowest glueball, is interpreted in~\cite{hooft} as the lowest $q\bar q$ scalar multiplet (see also~\cite{syracuse}).
The underlying hypothesis that the latter multiplet is not a radial excitation of the former has never been proven.
 
In this note we provide evidence in support of this hypothesis by estimating the mass  
of the first radial excitation of the lowest sub-GeV $\qq\bar\qq$ scalar meson.

We perform this calculation in  large-$N$ QCD in $(1+1)$-dimensions. This is a planar, 
linearly confining theory which admits a Bethe-Salpeter equation describing the discrete spectrum of $q\bar q$
bound states~\cite{coleman}. In this theory no orbital angular momentum excitations are possible since 
no rotation operator can be introduced: the discrete spectrum describes radial excitations.

Our Ansatz is that the scalar fields of  large $N$ 
scalar chromodynamics in $(1+1)$-dimensions (sQCD$_2$) can be thought as diquark fields. 
The corresponding Bethe-Salpeter equation of this theory should then 
yield the spectrum of tetraquarks $\qq\bar \qq$ particles.

The bound state equations in spinor and scalar chromodynamics are respectively (see e.g.~\cite{tsau}):
\begin{eqnarray}
&&\mu_{(\pi)}^2 \phi(x)=\frac{M_f^2}{x(1-x)} \phi(x)-\fint_0^1 dy \frac{1}{(y-x)^2}\phi(y)\label{uno}\\
&&\mu_{(\sigma)}^2\phi(x)=\frac{M_s^2-1}{x(1-x)}\phi(x) -\fint_0^1 dy \frac{1}{(y-x)^2}\frac{(x+y)(2-x-y)}{4x(1-x)}\phi(y)
\label{due}
\end{eqnarray}
where the integrals are in the sense of the Cauchy principal  value and $\phi(0)=\phi(1)=0$. 
Here $\mu^2_{(\pi)}$  and $\mu^2_{(\sigma)}$  are the mass squared eigenvalues of standard $q\bar q$ mesons, call them pions, and 
tetraquark  $\qq\bar\qq$ mesons, sigmas.  The parameters $M_f$ and $M_s$ 
represent the masses of the quark $q$ and of the scalar quark $\qq$ respectively. In this notation all the masses are 
adimensional parameters since we are rescaling them by the couplings $g^2$: in two-dimensional scalar and 
spinor chromodynamics the  couplings have dimension of mass.

We cannot expect that the mass spectra in (s)QCD$_2$ reproduce numerically  the physical values of the masses
of real pions and sigmas but we can assume that the regularities in the spectra  of this kind of hadron-string models 
resemble those in the physical ones. To begin, we require that the ratio between the ground states 
$\mu_{(\sigma)}/\mu_{(\pi)}$ corresponds to the ratio between the lowest lying tetraquark object, the 
$\sigma(500)$, and the lowest standard $q\bar q$ meson, the pion. 
We find that this is allowed by a family of values  of the parameters $(M_f,M_s)$.

The first observed meson with the same quantum numbers of the pion is the $\pi(1300)$~\cite{pdg} which is therefore
a good candidate for the first radial pion excitation.
Considering that all bound states are alternately even or odd under parity~\cite{hooft2} and that 
the ground state for a $\qq\bar\qq$ particle has positive parity, the bound state corresponding to $\pi(1300)$  is the 
second excited state in the spectrum $\mu_{(\pi)}$. Then a simple calculation 
allows to estimate the mass of the first expected physical radial excitation of the $\sigma(500)$ meson.

It turns out that, spanning the set of allowed masses $(M_f,M_s)$, the minimum value for
 the first radial excitation of the $\sigma(500)$ tetraquark is at about $2600$~MeV. This would predict 
 an extremely broad state hardly identifiable experimentally.
 Moreover this result underscores that there is no possibility that the above 1~GeV scalar mesons in the range 
$1350\div 1700$~MeV could be excitations of the below 1~GeV ones. This 
strengthens the hypothesis that in the latter mass region
there is space only for the  lowest $q\bar q$ multiplet (and  likely a glueball).

In the last section we extend the discussion to the case of heavy-light diquarks introduced to
describe the $X,Y,Z$ particles found by Belle and BaBar. In particular we will focus on the
charged $Z(4430)$ particle recently observed by Belle. As this is a spin one resonance, spin one diquarks have to be considered.
We find that the bound state equation for scalar chromodynamics can be exploited also for a discussion
of heavy-light axial diquarks.

We start by  illustrating the method used to solve Eqs.~(\ref{uno},\ref{due}) and the numerical results obtained.

{\bf \emph{Large $N$ scalar QCD$_2$}}.
We will briefly review scalar chromodynamics in (1+1)~dimensions 
and outline the derivation of Eq.~(\ref{due}). A thorough discussion of Eq.~(\ref{uno}) 
in spinor chromodynamics is found in~\cite{coleman}. Omitting color indices,
the scalar chromodynamics Lagrangian density in light-cone coordinates
\begin{equation}
x^{\pm}=\frac{x^0\pm x^1}{\sqrt{2}}
\end{equation}
reads:
\begin{equation}
{\cal L}=\partial_\mu\varphi^\dag\partial^\mu \varphi -m^2\varphi^\dag\varphi+\frac{1}{2}(\partial_-A_+)(\partial^+A^-)-\frac{g}{\sqrt{N}}A_+(\varphi^\dag\partial_-\varphi-(\partial_-\varphi^\dag)\varphi)
\label{lag}
\end{equation}
where the $\mu$ index is $\mu=+,-$, the gauge fields are in the adjoint representation of $SU(N)$  whereas the scalar fields are in the fundamental
one and the light-cone gauge $A^+=A_-=0$ has been imposed. 
In this gauge no self-couplings of the gauge fields or seagull terms exist. 
The theory is asymptotically free and linearly confining. To make the self energy smooth at large $N$, we introduce $1/\sqrt{N}$ at each vertex.

The emergence of a linearly confining potential can be observed by writing the equation of motion for the field $A_+$:
\begin{equation}
\partial_-^2A_+=-2\frac{g}{\sqrt{N}}j_-
\end{equation}
which admits the following solution (use  $\partial_-|x^-|={\rm sign} (x^-)$ and $\partial_- {\rm sign}( x^-)=2\delta(x^-)$):
\begin{equation}
A_+=-\frac{g}{\sqrt{N}}\int dy_+ |x_+ -y_+|j_-
\label{linear}
\end{equation}
in absence of background fields.
Plugging Eq.~(\ref{linear}) into Eq.~(\ref{lag}) it turns out  that there is a linear potential
between charges. By dimensional analysis, $g$ has dimensions of energy. 

We will consider this theory in the 't~Hooft limit: large $N$ and $g$  held fixed. 
As $N=3$ the color transformation property of a diquark (the scalar quark $\varphi_\alpha$) and of 
an antiquark are the same: diquarks are $qq$ states with attraction in the color ${\bf \bar 3}$ channel. 
Extending the color group to $SU(N)$, with $N>3$, this diquark-antiquark correspondence is lost.
Here we treat  the fields $\varphi_\alpha$ as the fields of  building block, pointlike, diquarks. As for color transformations, 
this is appropriate for $N=3$, whereas for larger $N$ it is an extrapolation.
Alternatively one could consider the Corrigan-Ramond large $N$ limit~\cite{cr}. 
In the latter case quarks and `larks' are introduced, transforming as the
${\bf N}$ and ${\bf N(N -1)/2}$ representations of $SU(N)$ respectively. Larks $\ell$ are 
represented by antisymmetric tensors $\ell_{\alpha\beta}=-\ell_{\beta\alpha}$ coinciding
with antiquarks if $N=3$. A theory of only larks is equivalent to QCD.  In the CR large $N$ limit,
a baryon is represented by $qq\bar \ell$ whereas there are no color singlets made up of 
three quarks in the standard large N.

In the large $N$ limit only planar diagrams are relevant and quark loops are suppressed. Hence gluon lines are impassable barriers 
as there are no gluon-gluon interactions and any gluon line crossing would violate planarity.
\newpage
Eq.~(\ref{due}) is derived from the Bethe-Salpeter equation
expressed by the following diagrammatic relation:
\vskip2.0truecm
\begin{center}
\SetScale{0.65}
\SetOffset(135,10)
\fcolorbox{white}{white}{
  \begin{picture}(297,91) (214,-165)
    \SetWidth{0.5}
 \Text(230,-77)[l]{$=$}
 \Text(200,-45)[l]{$p$}
  \Text(200,-95)[l]{$r-p$}
    \Text(327,-45)[l]{$p$}
      \Text(327,-95)[l]{$r-p$}
      \Text(315,-77)[l]{$k$}
    \GOval(227,-118)(12,13)(0){0.882}
 \Text(145,-77)[l]{$\psi$}
    \SetWidth{1.0}
    \ArrowLine(237,-109)(317,-74)
    \SetWidth{0.5}
    \GOval(420,-120)(12,13)(0){0.882}
    \Text(270,-78)[l]{$\psi$}
    \SetWidth{1.0}
    \ArrowLine(429,-110)(509,-75)
    \ArrowLine(511,-165)(430,-130)
    \ArrowLine(317,-163)(236,-128)
    \SetWidth{0.5}
    \Photon(469,-93)(469,-147){7.5}{3}
    \Vertex(495,-81){5.66}
    \Vertex(496,-159){5.66}
    \Vertex(298,-83){5.66}
    \Vertex(298,-155){5.66}
  \end{picture}
}
\end{center}
\vskip-2truecm
where the disks on the external legs indicate the dressed propagators. 
The shaded blobs represent the matrix element of the time ordered product of two scalar quark fields 
between the vacuum $|0\rangle$ and the meson state $|B\rangle$~\cite{coleman}.
We will denote by $\psi$ the Fourier transform of this matrix element.
The Bethe-Salpeter equation is the Dyson-Schwinger equation for a 4-point Green function $G$
where a bound state $|B\rangle$ occurs as a mass pole (the residue at the pole giving the bound state wave function) in $G$:
\begin{equation}
G\to \frac{|B\rangle\langle B|}{q^2-\mu^2+i\epsilon}.
\end{equation}
The simplicity of the previous diagrammatic equation is a consequence of the large $N$ limit.
Let us represent the Dyson-Schwinger series as:
\vskip2.3truecm
\begin{center}
\SetScale{0.65}
\SetOffset(125,10)
\fcolorbox{white}{white}{
  \begin{picture}(231,85) (221,-190)
    \SetWidth{0.5}
    \GOval(232,-146)(10,11)(0){0.882}
    \SetWidth{1.0}
    \ArrowLine(241,-138)(452,-110)
    \ArrowLine(452,-184)(240,-155)
    \SetWidth{0.5}
    \Vertex(435,-112){5.66}
    \Vertex(435,-182){5.66}
    \Text(290,-95)[l]{$...$}
    \GBox(377,-190)(404,-106){0.882}
    \GBox(284,-191)(311,-107){0.882}
  \end{picture}
}
\end{center}
\vskip-2.0truecm
In the large $N$ limit, the only non-vanishing contribution  from the amputated functions 
represented by the shaded boxes is the first one:
\vskip2.3truecm
\begin{center}
\SetScale{0.65}
\SetOffset(125,10)
\fcolorbox{white}{white}{
  \begin{picture}(255,86) (284,-189)
    \SetWidth{0.5}
    \GBox(284,-190)(311,-106){0.882}
    \SetWidth{1.0}
    \Photon(378,-109)(381,-188){5}{8}
    \ArrowLine(451,-109)(535,-109)
    \ArrowLine(532,-184)(453,-184)
    \Photon(452,-109)(533,-183){5}{8}
    \Photon(454,-184)(535,-108){5}{8}
    \Text(220,-95)[l]{$=$}
    \Text(270,-95)[l]{$+$}
      \Text(370,-95)[l]{$+$~~$...$}
  \end{picture}
}
\end{center}
\vskip-2.3truecm
Once the gluon propagator is found to be (see~\cite{coleman}):
\begin{equation}
D_{\mu\nu}=i\delta_{\mu+}\delta_{\nu+}\frac{\mathbb{P}}{k_-^2}
\end{equation}
where $\mathbb{P}$ indicates the Cauchy principal value, it is not difficult to compute the quark dressed propagators $D(p)$. 
This can bee seen by observing that:
\begin{equation}
\fint dk_- \frac{e^{ik_- x_+}}{k_-^2}=|x_+|
\end{equation}
\label{nome}
One obtains (cfr.~\cite{tsau}):
\begin{equation}
D(p)=\frac{i}{2p_-\left(p_+ -\frac{g^2}{2\pi}\left(\frac{{\rm sign}(p_-)}{\lambda}-\frac{1}{p_-}\right)+\frac{i\epsilon-M^2}{2p_-}\right)}
\label{propa}
\end{equation}
where $\lambda$ is an infrared cutoff. To derive the latter expression it is convenient to observe that:
\begin{equation}
\int dp_+ \frac{p_-}{2p_+p_--A+i\epsilon}=-i\frac{\pi}{2}{\rm sign}(p_-)
\end{equation}
as can be proved by using the Plemelij identity:
\begin{equation}
\int dx \frac{f(x)}{x-x_0\mp i\epsilon}=\fint dx \frac{f(x)}{(x-x_0)}\pm i\pi f(x_0)
\label{Plemelij}
\end{equation}
and that $\mathbb{P}/x=0$.

According to Eq.~(\ref{propa}), in the limit $\lambda\to 0$ the quarks have infinite self-energy, hence are removed from the spectrum. Anyway the dependency on $\lambda$ is canceled in  the Bethe-Salpeter equation which eventually defines the eigenvalue equation for scalar quark bound states; in other words  we have a discrete spectrum of bound states but no free quarks. The diagrammatic equation given above writes as:
\begin{equation}
\psi(p,r)=g^2 D(p)D(p-r)\fint\frac{d^2k}{4\pi^2}\frac{i}{k_-^2}\psi(p+k,r)(2p+k)_-(2p-2r+k)_-
\label{bse}
\end{equation}
Note the derivative couplings of scalar chromodynamics at the vertices of the gluon propagator. Observe also 
that the kernel of the principal value integral depends on minus momenta only: this is clear from Eq.~(\ref{linear}) where only $j_-$ currents are coupled to 
the linear potential, or equivalently from the expression of the gluon propagator. We can therefore  integrate both sides of Eq.~(\ref{bse}) 
over $p_+$. A straightforward residue calculation gives:
\begin{equation}
\int dp_+\frac{1}{(p_+-A)(p_+-r_+-B)}=-\frac{2\pi i\theta(p_-)\theta(r_--p_-)}{A-B-r_+}
\label{residue}
\end{equation}
where we required a negative imaginary part for the pole $A$ and a positive one for the pole $B$. 
If both poles were on the same half-plane the residue integration would yield zero closing the contour in the other half-plane. 
We close the contour of integration in the lower-half plane (clockwise sign). Defining:
\begin{equation}
\phi(p_-,r)\equiv\int dp_+\psi(p_+,p_-,r)
\end{equation}
one can write Eq.~(\ref{bse}) as:
\begin{equation}
\phi(p_-,r)=-\frac{g^2}{2\pi}\frac{\theta(p_-)\theta(r_- -p_- )}{r_+ - (A-B)}\int_0^{r_-}
\frac{dk_-}{(k_- - p_-)^2}\phi(k_-,r)\frac{(p_-+k_-)}{2p_-}\frac{(p_- + k_- -2 r_-)}{2(p_- -r_-)}
\label{bes2}
\end{equation}
The latter integral is infrared divergent when $p_-\sim k_-$. We can expand it as:
\begin{eqnarray}
\int_0^{r_-}\frac{dk_-}{(k_- - p_-)^2}\phi &=&\fint_0^{r_-}\frac{d k_-}{(k_- -p_-)^2}\phi-\phi(p_-,r)
\lim_{\lambda\to 0}\int_{p_--\lambda}^{p_-+\lambda}\frac{dk_-}{(k_- - p_-)^2}=\notag\\
&=&\fint_0^{r_-}\frac{d k_-}{(k_- -p_-)^2}\phi+\frac{2}{\lambda}\phi(p_-,r)
\end{eqnarray}
This expression allows to cancel the $1/\lambda$ infrared divergence in Eq.~(\ref{bes2}).
Consider in fact that:
\begin{equation}
A-B=\frac{g^2}{\pi}\frac{1}{\lambda}+\frac{M^2-g^2/\pi}{2p_-}-\frac{M^2-g^2/\pi}{2(p-r)_-}.
\end{equation}
Eq.~(\ref{due}) is obtained by defining $r_+=\mu^2/(2 r_-)$, $xr_-=p_- $ and $yr_-=k_-$ and rescaling all the square masses in units of $g^2/\pi$.
Observe that the $\theta$-functions in Eq.~(\ref{residue})  define an interval outside which $\phi=0$. This interval is $p_-\in [0,r_-]$ or $x\in [0,1 ]$.
In particular we have $\phi(0)=\phi(1)=0$.

The integral in Eq.~(\ref{due}) gives its main contribution if $y\simeq x$, where the kernels of Eq.~(\ref{uno}) and~(\ref{due}) are the same.  
As the highest part of the spectrum is concerned, one can neglect $M_{f,s}$ and the eigenfunctions are approximated by
$\phi(x)\simeq \sin \omega x= \sin n\pi x$, with $n$ integer $n>1$. This approximation respects the condition $\phi(0)=\phi(1)=0$.
For periodic functions $\phi(x)$ we have:
\begin{equation}
\fint_0^1dy \frac{\exp (i\omega x)}{(y-x)^2}\simeq \fint_{-\infty}^{\infty} dy\frac{\exp (i\omega x)}{(y-x)^2}=i\omega  \fint_{-\infty}^{\infty}dy \frac{\exp (i\omega x)}{(y-x)}= -\pi 
|\omega| \exp(i\omega x)
\end{equation}
as can be seen easily by applying~(\ref{Plemelij}). This result means that the eigenvalues in the highest part of the spectrum are $\mu^2_n\simeq n \pi^2$, 
i.e., there is {\it no continuum} in the spectrum~\cite{coleman}.  The alternating parity for these $\phi_n(x)$ states is evident.

In what follows we will focus on the lower part of the spectrum, and we will solve Eqs.~(\ref{uno}) and~(\ref{due}) with a numerical approach.

{\bf \emph{Discrete Spectra}}. 
Equations~(\ref{uno},\ref{due}) are integral equations with singular kernels and a prescription, 
the Cauchy principal value, on how to treat the singularity. In this section we will indicate the steps to put them in a
form amenable to numerical computation. We follow a procedure which has been applied in~\cite{ben} (and reference therein)
to solve Eq.~(\ref{uno}). The main point is to write:
\begin{equation}
\phi(\theta)=\sum_{m=0}^{N}a_m \sin m\theta
\label{fou}
\end{equation}
and set: 
\begin{eqnarray}
&&x=\frac{1+\cos\theta}{2}\\
&&y=\frac{1+\cos\theta^\prime}{2}
\end{eqnarray}
Integrating by parts in Eq.~(\ref{uno})  to lower the singularity of the kernel,  and exploiting the conditions $\phi(0)=\phi(1)=0$, one obtains 
an integral of the form:
\begin{equation}
\fint_0^\pi d\theta^\prime\frac{\cos m\theta^\prime}{\cos\theta^\prime-\cos\theta} 
\end{equation}
which can be solved using the following relation between Chebyshev polynomials~\cite{abramowitz}
\begin{equation}
\label{ceb1}
\fint_{-1}^1dy\frac{T_n(y)}{(y-x)\sqrt{1-y^2}}=\pi U_{n-1}(x),
\end{equation}
where $x\in[-1,1]$. Indeed this is done by observing that $T_n(\cos\theta)=\cos n\theta$ 
whereas $U_n(\cos\theta)=\sin(n+1)\theta/\sin\theta$. The next step is to discretize the angle $\theta$ according to 
the  prescription~\cite{ben}:
\begin{equation}
\theta\to\theta_k=\frac{k\pi}{N+1}
\end{equation}
and to use the following orthogonality relation:
\begin{equation}
\sum_{k=1}^{N}\sin k\theta_m\sin k\theta_n\equiv\sum_{k=1}^{N}\sin \theta_{k m}\sin \theta_{k n}=\frac{N+1}{2}\delta_{m n}
\end{equation}
to transform Eq.~(\ref{uno}) into an eigenvalue equation of the form:
\begin{equation}
\sum_{m=1}^{N}{\cal O}_{n m}^{(\pi)} a_m=\mu_{(\pi)}^{2}a_n
\label{eval1}
\end{equation}
where we have~\cite{ben}:
\begin{equation}
{\cal O}_{n m}^{(\pi)}=\frac{4}{N+1}\sum_{k=1}^{N}\frac{\sin\theta_{kn}\sin\theta_{k m}}{\sin\theta_k}\left(\frac{2M_f^2}{\sin\theta_k}+m\pi\right).
\end{equation}
The operator ${\cal O}_{nm}^{(\pi)}$ can be diagonalized and a set of discrete eigenvalues $\mu_{(\pi)}$ can be found. We aim here to find
an operator ${\cal O}_{nm}^{(\sigma)}$ and an eigenvalue equation form  for   Eq.~(\ref{due}) similar to that in~(\ref{eval1}).
The procedure to follow is basically 
the same as the one outlined above, just requiring some more algebra. An helpful relation to use to obtain ${\cal O}_{nm}^{(\sigma)}$ is provided by the following
identity relating Chebyshev polynomials, similar to that given in Eq.~{(\ref{ceb1})~\footnote{
The integrals we are concretely concerned with are particular cases of the integral relations between Chebyshev polynomials given in the text. In particular the two following integrals are needed:  
$\int_0^\pi d\phi \cos n\phi /(\cos \phi-\cos\theta)=\pi \sin n\theta/\sin\theta$ and $\int_0^\pi d\phi \sin n\phi\sin\phi/(\cos \phi-\cos\theta)=-\pi\cos n\theta$. An elegant derivation of these two integrals can be found in~K.~Karamcheti, {\it Principles of Ideal Fluid Aerodynamics}, Krieger (1980). See Appendix~E.}:
\begin{equation}
\fint_{-1}^1dy\frac{\sqrt{1-y^2}U_{n-1}(y)}{(y-x)}=-\pi T_n(x)
\label{ceb2}
\end{equation}
The result is:
\begin{eqnarray}
{\cal O}_{n m}^{(\sigma)}&=&\frac{2}{N+1}\sum_{k=1}^{N} \left[ \left(\frac{4(M_s^2-1)}{\sin^2\theta_k}+\frac{7}{16}m\pi\frac{1}{\sin\theta_k}\right)\sin\theta_{nk}\sin\theta_{mk}-\frac{\pi}{2}\cos\theta_k\sin\theta_{nk}\cos\theta_{mk}+\right.\notag\\
&&\left.-\frac{m\pi}{8}\cot\theta_k\sin\theta_{nk}\left(  \sin\theta_{mk}\cos\theta_k +\sin\theta_{(m+1)k}+\sin\theta_{(m-1)k} \right)+\right.\notag\\
&&\left. -\frac{m\pi}{32}\frac{\sin\theta_{nk}}{\sin\theta_k}\left( \sin\theta_{(m+2)k}+\sin\theta_{(m-2)k}\right)  \right].
\end{eqnarray}
In the derivation of ${\cal O}_{n m}^{(\sigma)}$ the trigonometric Werner identities have been used. 
Thus we have to pick up a minimum value of $M_s$ to avoid tachyons.

The operator obtained allows to write also Eq.~(\ref{due}) in a form amenable to numerical computation:
\begin{equation}
\sum_{m=1}^{N}{\cal O}_{n m}^{(\sigma)} a_m=\mu_{(\sigma)}^2a_n
\end{equation}
and can be diagonalized to obtain the spectrum of eigenvalues $\mu_{(\sigma)}$.

{\bf \emph{Numerical results}}.  In Figure~\ref{figu} we show a set of  values $(M_s,M_f)$ which give a ratio $\mu_{(\sigma)}/\mu_{(\pi)}$ compatible with the 
experimental one, taking the mass value of  the $\sigma$ from~\cite{hooft}.  
\begin{figure}[htb]
\begin{center}
\epsfig{file=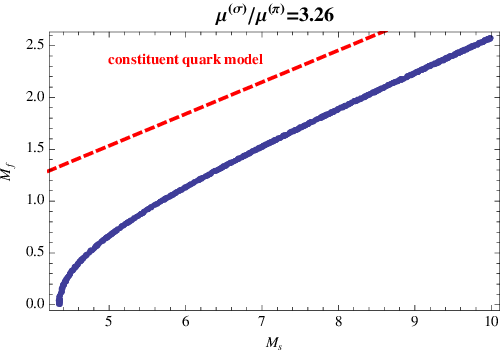,width=6cm}
\caption{ The curve represents the $(M_s,M_f)$ values which give the ratio $\mu_{(\sigma)/}\mu_{(\pi)}$ of the lowest eigenvalues 
equal to the experimental one. We check that for large values of 
$(M_f, M_s)$ the two equations~(\ref{uno})~and~(\ref{due}) are the same and the results converge with what expected from naive constituent quark model. The numerical analysis is performed by setting the $N$ parameter in Eq.~(\ref{fou}) to $N=100$. The results are stable under variation of $N$.}
\label{figu}
\end{center}
\end{figure}
These values can be used to determine the gap between the ground levels and the first excited levels, having the desired parity, in scalar  and spinor chromodynamics. The first radially excited $0^-$ state is expected to be the second excited state in the $q\bar q$ spectrum, the ground state being $0^-$. 
Define the following ratio:
\begin{equation}
\lambda=\left(\frac{\mu_{(\pi)}^{[2]} }{\mu_{(\pi)}^{[0]}}\right) /\left(  \frac{\mu_{(\sigma)}^{[2]} }{\mu_{(\sigma)}^{[0]}}\right)
\end{equation}
where with the superscripts $0,1,2,...$ we label the levels in the spectrum: $0$ is the ground state.
The predicted value for the first positive parity radial excitation of a system of two scalar quarks (a diquark-antidiquark state in our language) is then $m^{[2]}_{(\sigma)}$ 
which can be obtained by:
\begin{equation}
m^{[2]}_{(\sigma)}=\frac{m_{\pi^\prime}}{m_\pi}\frac{1}{\lambda}
\end{equation}
using the mass $m_{\pi^\prime}=1300$~MeV for the first radial excitation of the pion and $m_\pi=135$~MeV, $m_\sigma=470$~MeV.
With the values  $(M_s,M_f)$ shown in Fig.~\ref{figu} it turns out that $m^{[2]}_{(\sigma)}\gtrsim 2590$~MeV indicating that 
a radial excitation of the lowest lying tetraquark state would be likely a very broad object. 
Also there is no harm to find an exotic radial excitation in the range of masses where  the super-GeV scalar nonet is located.
Similarly we can also estimate the lower bound for the first radial $0^-$ excitation of a $\qq\bar\qq $ particle, finding $m^{[1]}_{(\sigma)}\gtrsim 3060$~MeV. This can be done by assuming that the first $q\bar q$ scalar radial excitation is the $f_0(1370)$. The given minimum value occurs $e.g.$ at the  point  $(M_f,M_s)\sim (0.1,4.3)$ of the parameter space spanned by the curve in Fig.~\ref{figu}.

{\bf \emph{Heavy-light diquarks. }}
Diquarks have been used in the literature also to discuss a number of recently discovered charmonium-like resonances, 
known as $X,Y,Z$ particles, with phenomenological properties conflicting with  those of standard charmonium  states. 
The $X(3872)$ found by Belle~\cite{belle},  was recognized from the start to be a very 
anomalous charmonium state, despite its discovery decay mode into $J/\psi \rho$ that could have seemed the footprint of a higher charmonium.
In~\cite{xnoi} the $X(3872)$ is interpreted as a $\qq\bar\qq$ state with $\qq=[cq]$, $q$ being a light quark $q=u,d$. The diquark model
predicts two charged almost degenerate $X$ particles and two neutral ones, namely $[cu][\bar c \bar u]$ and $[cd][\bar c \bar d]$. As 
discussed in~\cite{split} there is strong evidence by BaBar and Belle that a second neutral state, the $X(3876)$, exists! As for now, there are no
conclusive indications for $X^{\pm}$. On the other hand Belle has found three charged resonances decaying into charmonium+charged~$\pi$, namely
$Z_1(4051), Z_2(4248), Z(4430)$. These are produced in $B\to K Z_1$, $Z_2$ or $Z$.
One can find a brief account on this in~\cite{bellepr}. 

For example, the $Z(4430)$, observed in $Z\to \psi(2S)\pi^+$~\cite{bellez}, if confirmed, should necessarily be a multiquark object.
In~\cite{znoi}  this state is interpreted as a $[cu]_1[\bar c \bar d]_0$ radial excitation of a $1^{+-}$ particle predicted in~\cite{xnoi} and decaying
into $\psi(1S)\pi^+$.

Interestingly the difference in mass between the radially excited tetraquark $Z^+$ and its fundamental state predicted in~\cite{xnoi} is 
very close to $M_{\psi(2S)}-M_{\psi(1S)}$. 

The notation $[cu]_1[\bar c \bar d]_0$ means that we assume one of the two diquarks to have spin one.
Light diquarks with spin one are thought to be less stable than spin zero ones, but, as a heavy-quark is introduced in the diquark, 
spin-spin chromomagnetic 
interactions are suppressed by the heavy mass and one expects to treat heavy-light diquarks with spin zero and spin one on the same footing. 
For a phenomenological investigation on higher light tetraquarks see~\cite{drenska}.

In this section we want to investigate the system $\qq_1\bar\qq+ \qq\bar\qq_1$ using the methods developed above. 
We assume that the spin one diquark can be described by a colored (axial) field $\aa_i$ in the ${\bf \bar 3}$-color representation. 
The gluon field in the adjoint is $A^i_j$, with $A^i_i=0$ and $A^i_j=-A^{*j}_i$ if the minimal coupling is 
$\partial_\mu\to\partial_\mu+gA_\mu$.  We define ${\cal F}_{\mu\nu }^i=(\partial_\mu \aa_\nu^i-\partial_\nu \aa_\mu^i)$ for the axial diquark.

The vertex $\aa\aa A$ can be extracted by:
\begin{equation}
\frac{g}{\sqrt{N}}[\aa_i^\mu \aa^{\nu j} (F_{\mu\nu})^i_j+\aa^\mu_i {\cal F}_{\mu\nu}^j(A^{\nu})^i_j+{\cal F}_{\mu\nu i}\aa^{\mu j} (A^\nu )^i_j ]
\label{feyr}
\end{equation}
or in terms of  Feynman rules:
\vskip2.3truecm
\begin{center}
\SetScale{0.65}
\SetOffset(-15,10)
\fcolorbox{white}{white}{
  \begin{picture}(118,78) (287,-194)
    \SetWidth{1.0}
    \Photon(344,-116)(346,-191){7.5}{3}
    \ArrowLine(405,-191)(287,-191)
     \Text(280,-110)[l]{$=$}
      \Text(200,-115)[l]{$i$}
      \Text(250,-115)[l]{$j$}
       \Text(165,-135)[l]{$p-r+k,\mu$}
      \Text(235,-135)[l]{$p-r,\nu$}
      \Text(235,-85)[l]{$k,\alpha$}
       \Text(300,-110)[l]{$\frac{g}{\sqrt{N}}[ g_{\alpha\nu}(k-p+r)_\mu  +g_{\alpha\mu}(-p+r-2k)_\nu+g_{\mu\nu} (2p-2r+k)_\alpha]$}
  \end{picture}
}
\end{center}
\vskip-2.3truecm
where $\alpha,\mu, \nu=+,-$. The light cone gauge fixes $\alpha=-$ and $A=(\epsilon,-\epsilon)$. If the incoming $\aa$ is $\aa=(0,\epsilon)$ then the outgoing one is $\aa=(\epsilon,0)$, {\it i.e.},
at the vertices we have $\aa_+\aa_-$ or vice-versa. We sum over these two alternatives. Taking $\alpha=-, \nu=-, \mu=+$, from (\ref{feyr}) we get 
$(-p+r-2k)_-+(2p-2r+k)_-=(p-r-k)_-$ whereas taking $\alpha=+, \nu=+, \mu=-$ we have $(k-p+r)_-+(2p-2r+k)_-=(p-r+2k)_-$. The sum of the two contributions involves 
$(2p-2r+k)_-$ as in (\ref{bse}). The same considerations are applied to the calculation of the self-energy ($\aa A \aa$) which therefore proceeds in 
the same way as in the scalar chromodynamics case. 

This allows to use the same Bethe-Salpeter equation given in (\ref{due}) to study the case of charmed diquark-antidiquark states of the 
form $[cu]_1[\bar c \bar d]_0+[cu]_0[\bar c \bar d]_1$.

In~\cite{xnoi}, using chromomagnetic interaction Hamiltonians, the following mass values of $1^{+-}$, $[cq][\bar c \bar q]$ states, where found: $Z(3754)$ and
$Z(3882)$. In~\cite{znoi}  the hypothesis is made that the observed $Z^+(4430)$ could be a radial excitation of the $Z(3882)$. Here we assume that these 
three states, $Z(3754), Z(3882), Z(4430)$ correspond to $\mu^{[0]}, \mu^{[2]}, \mu^{[4]}$ of Eq.~(\ref{due}). If we require that $\mu^{[2]}/\mu^{[0]}=1.08=M_{Z(3882)}/M_{Z(3754)}$ we see that the minimum allowed value of the parameter $M_s$ (we could call it $M_a$ in this case making reference to axial diquarks) is $M_s\sim 10.43$, quite higher than what found before, consistently with the idea that we should describe spectra of charmed ({\it i.e.} heavier) particles. With the latter value of $M_s$ one finds
$\mu^{[4]}/\mu^{[2]}\sim1.20$ which requires that the value of the mass of $Z(4430)$, as found in this model, is indeed $4658$~MeV. This value is rather stable if one 
increases the parameter $M_s$ up to $M_s\sim 20$.  The next radial excitation is predicted at $6055$~MeV. 

{\bf \emph{Conclusions}}.
In this short note we assume that scalar chromodynamics in (1+1)-dimensions and with a large number of colors could be used to estimate the mass of the first radial excitation of a diquark-antidiquark meson. 
We applied a numerical method to solve the Bethe-Salpeter equations and compute the bound state discrete spectrum of this confining theory. The possible masses of the spinor and scalar quarks are found by imposing 
that the ratio of the ground state eigenvalues of the spinor and scalar Bethe-Salpeter equations,~Eqs.~(\ref{uno}) and~(\ref{due}) respectively, is equal to
the ratio of the physical masses $m_\pi/m_\sigma$. 
Furthermore, with these masses, we were able to extract a minimum value for the first radial
excitation of the ground state diquark-antidiquark spectrum. We extend our discussion to the heavy-light diquark sector finding that the 
$Z^+(4430)$ observed by Belle could correspond to the second radial excitation of the spectrum $Z(3754), Z(3882), Z(4658) ...$ where the lower states were 
predicted in~\cite{xnoi}.

\begin{acknowledgments}
We wish to thank M Bochicchio,  G D'Ambrosio, G Isidori, and J Schechter for very useful comments on the manuscript. 
ADP also thanks  R Escribano, R Faccini and N Drenska for several discussions and suggestions.
The work of one of us (BG) is  supported in part by the US Department of Energy under contract
DE-FG03-97ER40546.
\end{acknowledgments}

\bigskip 

\end{document}